\documentclass[sigconf]{acmart}

\usepackage[utf8]{inputenc}

\AtBeginDocument{
  \providecommand\BibTeX{{
    \normalfont B\kern-0.5em{\scshape i\kern-0.25em b}\kern-0.8em\TeX}}}

\copyrightyear{2020}
\acmYear{2020}
\setcopyright{acmlicensed}\acmConference[ETRA '20 Full Papers]{Symposium
on Eye Tracking Research and Applications}{June 2--5, 2020}{Stuttgart,
Germany}
\acmBooktitle{Symposium on Eye Tracking Research and Applications (ETRA '20
Full Papers), June 2--5, 2020, Stuttgart, Germany}
\acmPrice{15.00}
\acmDOI{10.1145/3379155.3391321}
\acmISBN{978-1-4503-7133-9/20/06}

\acmSubmissionID{1052}

\setcitestyle{square}

\usepackage{balance}
\usepackage{graphics}
\usepackage{booktabs}
\usepackage{soul}
\usepackage{enumitem}

\usepackage{ccicons}
\usepackage{xspace}
\usepackage{multirow}
\usepackage{tabularx}
\usepackage{caption}

\usepackage{wrapfig}
\usepackage{subcaption}
\newcommand{\ie}{\emph{i.e.\ }}
\newcommand{\eg}{\emph{e.g.\ }}

\newcommand{\twovotes}{\texttt{Vote2Metrics}\xspace}
\newcommand{\threestages}{\texttt{Vote3Stages}\xspace}

\def\plaintitle{Detecting Relevance during Decision-Making from Eye Movements for UI Adaptation}

\def\plainkeywords{UI adaptation; relevance detection; decision-making; eye tracking}

\begin{document}

\title{\plaintitle}

\author{Anna Maria Feit}\authornote{These authors contributed equally to this work.}
\affiliation{
  \institution{ETH Zurich}
}
\author{Lukas Vordemann}\authornotemark[1]
\affiliation{
  \institution{ETH Zurich}
}
\author{Seonwook Park}
\affiliation{
  \institution{ETH Zurich}
}
\author{Caterina Bérubé}
\affiliation{
  \institution{ETH Zurich}
}
\author{Otmar Hilliges}
\affiliation{
  \institution{ETH Zurich}
}
\setlength{\textfloatsep}{16pt}

\renewcommand{\shortauthors}{A. Feit et al.}

\begin{abstract}
This paper proposes an approach to detect information relevance during decision-making from eye movements in order to enable user interface adaptation. This is a challenging task because gaze behavior varies greatly across individual users and tasks and ground-truth data is difficult to obtain.
Thus, prior work has mostly focused on simpler target-search tasks or on establishing general interest, where gaze behavior is less complex.
From the literature, we identify six metrics that capture different aspects of the gaze behavior during decision-making and combine them in a voting scheme.
We empirically show, that this accounts for the large variations in gaze behavior and out-performs standalone metrics. Importantly, it offers an intuitive way to control the amount of detected information, which is crucial for different UI adaptation schemes to succeed.
We show the applicability of our approach by developing a room-search application that changes the visual saliency of content detected as relevant.
In an empirical study, we show that it detects up to 97\% of relevant elements with respect to user self-reporting, which allows us to meaningfully adapt the interface, as confirmed by participants.
Our approach is fast, does not need any explicit user input and can be applied independent of task and user.

 \end{abstract}



\maketitle

\section{Introduction}

Many decisions we make throughout the day are based on the information provided by a user interface (UI). For example, deciding which house appliance to order online, which job to apply for, or selecting a location for a hike. Designing UIs that support decision-making can be difficult. Showing all information  might not be useful due to device constraints~\cite{kajan2016} or to avoid information overload~\cite{Hwang1999}. Showing the right information is crucial for the decision quality~\cite{Papismedov2019}, but identifying it can be hard since users perceive the relevance of information differently (\eg~\cite{orquin2013}), an aspect that cannot be foreseen at design time. Thus there arises a need for adaptive UIs that understand which parts are relevant for a user, ideally without having to interrupt them.

In this paper, we propose an eye-tracking based approach to detect the relevance of displayed information while a user is making a decision. We demonstrate that it can be flexibly used for different UI adaptation schemes (such as the emphasis or suppression of relevant or irrelevant information) across different application domains without requiring any explicit user input.
While directly asking the user is a good way to understand which part of a UI is relevant to them, this compromises user experience and can be unreliable~\cite{aribarg2010,lieberman2001}. Eye tracking has proven to be an objective measure for a person's attention during visual search~\cite{duchowski2002,salojaervi2005}. However, gaze behavior during decision-making is even more user- and context-dependent~\cite{orquin2013}. For instance, it can vary drastically across users who employ different decision strategies (see \cite{orquin2013, gigerenzer2011, payne1992constructive, simon1957}), but also for the same user as they transitions from obtaining an overview of the UI to comparing relevant information to finally validating their decision~\cite{gidlof2013,russo1994}.
This makes detecting information relevance during decision-making a challenging task and prior work mostly considers simpler visual search tasks.

The goodness of a relevance detector highly depends on the UI adaptation scheme it is applied to. In particular, it is important to minimize the risk of usability issues due to incorrect adaptations~\cite{Findlater2009}. For example, wrongly highlighting irrelevant elements in a menu can induce a performance cost~\cite{Findlater_ephemeral_2009}.
In such a case, a successful relevance detector should  minimize  false positives (content wrongly predicted as relevant). On the other hand, when removing irrelevant elements, we need to prioritize the maximization of true positives (content correctly identified as relevant) to not hide any relevant information.
Ultimately, a good relevance detector should not reduce usability or induce any cognitive dissonance between the adapted UI, and what the user expects to see.

Embracing these requirements, we propose an approach to detect information relevance during decision-making from user eye movements.
We build on prior work that showed the efficacy of simple gaze metrics (\eg fixation duration and overall dwell time) in inferring covert attention (the mental focus of a person) from user gaze behavior \cite{kajan2016,Faro2010,alt2012}.
We select six well-established gaze metrics that capture different aspects of the gaze behavior during decision-making and can thus account for variations in decision strategies across users. Each metric can be seen as a weak relevance detector which makes a binary decision on whether an element is relevant or not. These decisions are considered as individual votes which are summed and thresholded to yield a final decision. This is motivated by the machine learning literature on multiple classifier systems~\cite{polikar2006ensemble} and, loosely speaking, has similarity to ensemble methods.

Through varying the threshold of votes we obtain a set of recognizers that trade-off true positives and false positives in different ways. This allows us to easily choose the right recognizer for a specific UI adaptation scheme. We validate this on a pilot dataset of stock-trading decisions and choose two thresholds that yield different trade-offs. We develop a proof-of-concept application of a realistic apartment share website and apply the resulting recognizers to visually highlight relevant information or fade out irrelevant ones, based on the  gaze data of user deciding whether they should apply for a room.
In an empirical study with 19 participants, all but one stated that the adaptation correctly highlighted relevant information and faded the irrelevant ones, without experiencing any usability issues. We find that the two applied voting schemes correctly recognize 97\%  and 80\% of elements reported as relevant by participants, at a false positive rate of 42\% and 17\% respectively.

In summary, we propose an unobtrusive approach to detect information relevance from eye movements during decision-making. It combines six gaze metrics in a voting scheme which captures the different gaze characteristics during decision-making. Varying the number of votes yields different recognizers that are effective in  emphasizing or suppressing information in an adaptive UI.
Our method is simple to understand and can be easily implemented, for which we will provide an open-source implementation.

 \section{Background \& Related Work}

To detect the relevance of information or infer user's interest or intent, researchers have explored the use of implicit feedback mechanisms, such as mouse and keyboard interactions~\cite{Jayarathna2015, Kelly01}. In this line of work, eye tracking has proven to be an unobtrusive and objective measure for a person's attention~\cite{duchowski2002,salojaervi2005} and was suggested to be more reliable than feedback from mouse movements, clicks, or scrolling~\cite{buscher2012wsdm,kelly2003,navalpakkam2013}.
However, it is a major challenge to infer \emph{covert} attention (\ie the mental focus of a person) from eye movements and relate it to the relevance of the displayed information. One reason is that eye gaze is not always strategically controlled but also stimulus-driven~\cite{borji2013}, \ie attracted by visually salient features of a UI.
Therefore, researchers have proposed various \emph{gaze metrics} to aggregate the noisy eye movements and analyze the underlying cognitive processes (\eg see ~\cite{jacob2003eye} for an overview).
In particular, the \emph{sum of fixations} (the number of times the user focuses an area of interest (AOI)) and the \emph{overall dwell time} (time spent looking at an AOI) have been successfully used to infer users' interest.
However, most of the prior work focused on well-defined tasks, such as visual target-search, where the ground truth is well-defined  (\eg~\cite{Aula2005,Dumais2010,gwizdka2014,kajan2016,salojaervi2003}).
Only few researchers have considered more unconstrained settings where ground truth labels are not available, mostly to establish general interest in displayed information~\cite{alt2012,Nguyen2016,qvarfordt2005}. These combine few well-established gaze metrics but are highly tuned to their specific applications.

This paper is one of few that tackles the more complex task of analyzing gaze behavior during decision-making. In the following, we give a brief introduction to the decision-making literature that informed our approach and review prior work that used eye movements to adapt a UI.

\subsection{Gaze behavior during decision-making}
Decision-making is a complex cognitive process, during which eye movements play an important role in order to retrieve displayed information, \eg to encode and process it for the first time or to update a person's working-memory~\cite{orquin2013}.
However,  the relevance of information varies even during the decision process of an individual. We refer to \citet{orquin2013} for an extensive review of eye movements during decision-making and here summarize the phenomena fundamental to our work.

\subsubsection{Decision strategies}
The theory of bounded rationality suggests that people prioritize information according to their individual \emph{decision strategy} or heuristics~\cite{simon1957}, as a consequence of their cognitive capacity limitations (for a review see \eg~\cite{gigerenzer2011, orquin2013}). Accordingly, gaze behavior varies according to a person's decision strategy. At the same time, learning influences the eye movements, in that repeated exposure to the same decision decreases the time it takes to fixate relevant elements, while the duration of fixations increases~\cite{jovancevic2009}.
Characteristic for the gaze behavior during decision-making, are strategic just-in-time return-fixations on relevant information in order to reduce the working-memory load. The specific trade-off between retrieving information through fixations or from working memory varies across users and tasks~\cite{Droll2007}.
Our selection of gaze metrics is motivated by these findings.

\subsubsection{Cognitive stages during decision-making}
Several researchers have argued that the decision-making process can be separated into three cognitive stages which are characterized by different eye movements~\cite{gidlof2013,russo1994}. Following \citet{russo1994}, we refer to them as (1) Orientation, (2) Evaluation, and (3) Verification.
In the orientation stage the user obtains an overview of the available options or information. It is characterized by a scanning pattern with a series of shorter fixations without any return-fixations on already encoded information.
During the evaluation stage, the user compares the information determined as relevant during orientation, which is marked by many return-fixations. The final verification stage again uses short fixations on the most relevant information to validate the decision (see ~\cite{orquin2013}).
An exact separation of these stages based on changes in gaze behavior is difficult and disputed in the literature~\cite{ gidlof2013,russo1994}.
However, we can use them as a theoretical motivation for choosing gaze metrics that each capture a different aspect of the decision process in order to obtain a more holistic view of a person's gaze behavior than overall dwell time or number of fixations alone could provide.

\begin{table*}[t]
\captionsetup{font=small}
    \caption{Our selection of gaze-metrics and their associated stages during decision making.}
    \vskip -3mm
    \setlength{\tabcolsep}{4pt}
    \renewcommand{\arraystretch}{0.9}
    \sffamily
    \small
    \begin{tabular}{p{0.02\textwidth}p{0.13\textwidth}p{0.5\textwidth}p{0.24\textwidth}}
    \hline
    \multicolumn{4}{c}{\textbf{Orientation}} \\
    \hline
    TFF & \emph{Time to First Fixation} &  The time elapsed between the presentation of a stimulus and the first time that gaze enters a given AOI. A low TFF value indicates high relevance. & \cite{byrne1999eye,gegenfurtner2011expertise}.\\
    FPG & \emph{First Pass Gaze} &  The sum of duration of fixations on an AOI during the first pass, \ie when the gaze first enters and leaves the AOI. A high FPG value indicates high relevance. & \cite{henderson1998,salojaervi2005}\\
    \hline
    \multicolumn{4}{c}{\textbf{Evaluation}} \\
    \hline
    SPG & \emph{Second Pass Gaze} & The sum of duration of fixations on an AOI during the second pass. A high SPG value indicates high relevance. & \cite{henderson1998}  \\
    RFX & \emph{Refixations Count} & The number of times an AOI is revisited after it is first looked at. A high RFX value indicates high relevance. & \cite{kajan2016,klami2010,Nguyen2016,salojaervi2005} \\
    \hline
    \multicolumn{4}{c}{\textbf{Verification}} \\
    \hline
    SFX & \emph{Sum of Fixations} &
        The total number of fixations within an AOI. A high SFX value indicates high relevance. & \cite{balatsoukas2012,kajan2016,klami2010} \\
    ODT & \emph{Overall Dwell Time} &
        The total time spent looking at an AOI including fixations and saccades. A high ODT value indicates high relevance. & \cite{alt2012,bednarik2012you,Nguyen2016} \\
    \hline
    \end{tabular}
    \label{tab:metrics_list}
\end{table*}
\subsection{UI adaptation from gaze behavior}
Eye gaze has been recognized as an unobtrusive feedback channel that provides rich information about a user's attention~\cite{qvarfordt2005}. This opened opportunities for adaptive user interfaces.
A large body of work has focused on enhancing the query-based search of images~\cite{Faro2010, klami2008, klami2010} or text-documents~\cite{Aula2005,buscher2008,Dumais2010, Bhattacharya2020}. There, information about eye gaze provides feedback on the relevance of the displayed search results or text documents.
Similarly, advertisements on a website were adapted based on relevance inferred from overall dwell time~\cite{alt2012}.
Several works have analyzed the gaze behavior while exploring a map~\cite{krejtz2017}, for example to adapt complex legends to only show relevant items~\cite{gobel2018}, to highlight important points of interest to facilitate planning~\cite{gobel2019}, or to follow-up on user's interests~\cite{qvarfordt2005}.
Other systems extract relevant information to compile a summary for later use~\cite{buscher2012tiis, Nguyen2016}.
UI adaptation based on user's cognitive load or context has recently been shown for mixed-reality settings~\cite{Gebhardt2019,lindlbauer2019} where the design of a good UI can be difficult as the user's context frequently changes, similar to other applications of ubiquitous computing~\cite{Dourish2004}.

Most of these work detect relevance by using well-established gaze metrics often in a way that is highly tuned to their specific application. The most related work to ours is \citet{Gebhardt2019}, where they detect relevance during purchasing decisions in an unsupervised fashion using a reinforcement learning approach. Their method is highly task specific and requires retraining for each new task. In the following, we show that our relevance detector is successful across the very different decision tasks of financial trading and room search, and works without any prior training.

\section{Successful relevance detection for adapting a User Interface}
The goal of our work is to detect the relevance of information displayed by a user interface (\ie of UI elements) in order to enable adaptation of the UI. The adaptation techniques considered here can, broadly speaking, be divided into two types. They either
\begin{enumerate}[leftmargin=*]
    \item \emph{emphasize relevant content}, \eg through coloring, rearranging, or replicating UI elements; or
    \item \emph{suppress irrelevant information}, \eg by greying out, removing, or hiding elements in a different part of the UI (see \eg~\cite{Deuschel2016, Findlater2009}).
\end{enumerate}
Correspondingly, a successful recognizer in the first case identifies a subset of relevant UI elements (true positives), whereas in the second, we are interested in identifying non-relevant ones (true negatives). At the same time, it is important to consider the consequences of misrecognitions: if incorrect adaptations lead to usability issues or high interaction costs, the adaptive interface is not successful~\cite{Findlater2009}.

In the first case, emphasizing content that is not considered as relevant by the user (false positives) could result in cognitive dissonance. For example, highlighting irrelevant information through coloring deflects attention and makes it harder for the user to focus on relevant content. In contrast, failing to emphasize an element regarded as relevant by the user (false negatives) has a comparably low cost, since its appearance does not change in comparison to the non-adapted version. As a result, a successful recognizer for emphasizing information should achieve a low false positive rate.

In the second case of suppressing less relevant information, an incorrect adaptation can hide information that is regarded as relevant (false negatives). This can lead to high interaction costs in order to recover such information, if even possible. In contrast, failing to hide an element because it is wrongly recognized as relevant (false positives) does not change the user interface in comparison to a non-adapted version. As a result, a successful recognizer for suppressing information should achieve a high true positive rate.

In conclusion, depending on the adaptation scheme, emphasizing or suppressing information, a recognizer has to trade-off true positives and false positives differently. In the following we present our voting scheme and show that it allows us to flexibly choose a good trade-off by simply adjusting the number of votes required to recognize a UI element as relevant.

\section{Method}

As discussed above, gaze behavior during decision-making varies significantly across people depending on their decision strategy. At the same time, the eye movements of a single user vary as she goes through different stages of decision-making (orientation, evaluation, an verification). As a consequence, often-used gaze metrics, such as overall dwell time or number of fixations alone cannot suffice to capture the relevance of each UI element, hereafter called area of interest (AOI). For example, when comparing two AOIs during the evaluation phase, the gaze frequently returns to the same AOI albeit for a short amount of time. The refixation count is a good indicator of its relevance, which would not be captured by the overall dwell time. To account for these variations, we select six well-established gaze metrics from the literature (see Table~\ref{tab:metrics_list}) that are able to capture the relevance of elements from the different gaze characteristics at each decision-making phase. We propose to combine them in a voting scheme with which one can flexibly trade-off true positives and false positives by changing the minimum number of metrics required to indicate relevance of the AOI. This is shown based on a pilot dataset of stock investment decisions.

\subsection{Gaze metrics}

Table~\ref{tab:metrics_list} shows the six gaze metrics we selected to capture different characteristics of the gaze behavior associated with the three stages of decision-making.
It includes metrics that capture the relevance of AOIs for establishing an initial overview of available information; metrics that capture the in-depth analysis of relevant information; and metrics that summarize the overall attention given to an AOI.

\subsection{Voting approach}
Our proposed voting scheme
considers each metric as informing a single-expert-classifier. For instance, by comparing ODT values across $n$ AOIs in an interface, we can determine if the value of a specific AOI is comparably high, indicating its relevance. By allowing multiple metrics to vote on an AOI's relevance, we imitate a multiple-classifier system while avoiding the need for training data.

We say that a metric casts a vote for a given AOI as being relevant, if its standard score (z-score) for the AOI is positive. Intuitively, this means that for the given AOI, the gaze characteristic captured by the metric deviates from the average across all AOIs. Note, that in the case of TFF, a negative z-score indicates relevance (shorter time to first fixation). The z-score is computed per stimuli (\ie for each decision made with a UI). If gaze data of several stimuli is available, the z-scores are averaged before a vote is cast.
To finally detect the relevance of an AOI, we count the number of votes cast by the $6$ metrics and compare it to a threshold. Requiring a higher number of votes would yield a lower number of AOIs being classified as relevant.
In the next section, we exhaustively analyze the effect of this threshold on the true and false positive rate of the voting scheme based on a pilot dataset of financial trading decisions.

In contrast to prior work, our voting approach requires no ground-truth data from a target person in order to tune thresholds for combining different metrics~\cite{kozma2009gazir,klami2010}. Moreover, it assumes no fixed number of relevant AOIs~\cite{kozma2009gazir,klami2010} but can flexibly select a varying number of AOIs.

\begin{figure}[t]
\setlength\belowcaptionskip{-10pt}
\centering
\includegraphics[width=0.7\columnwidth]{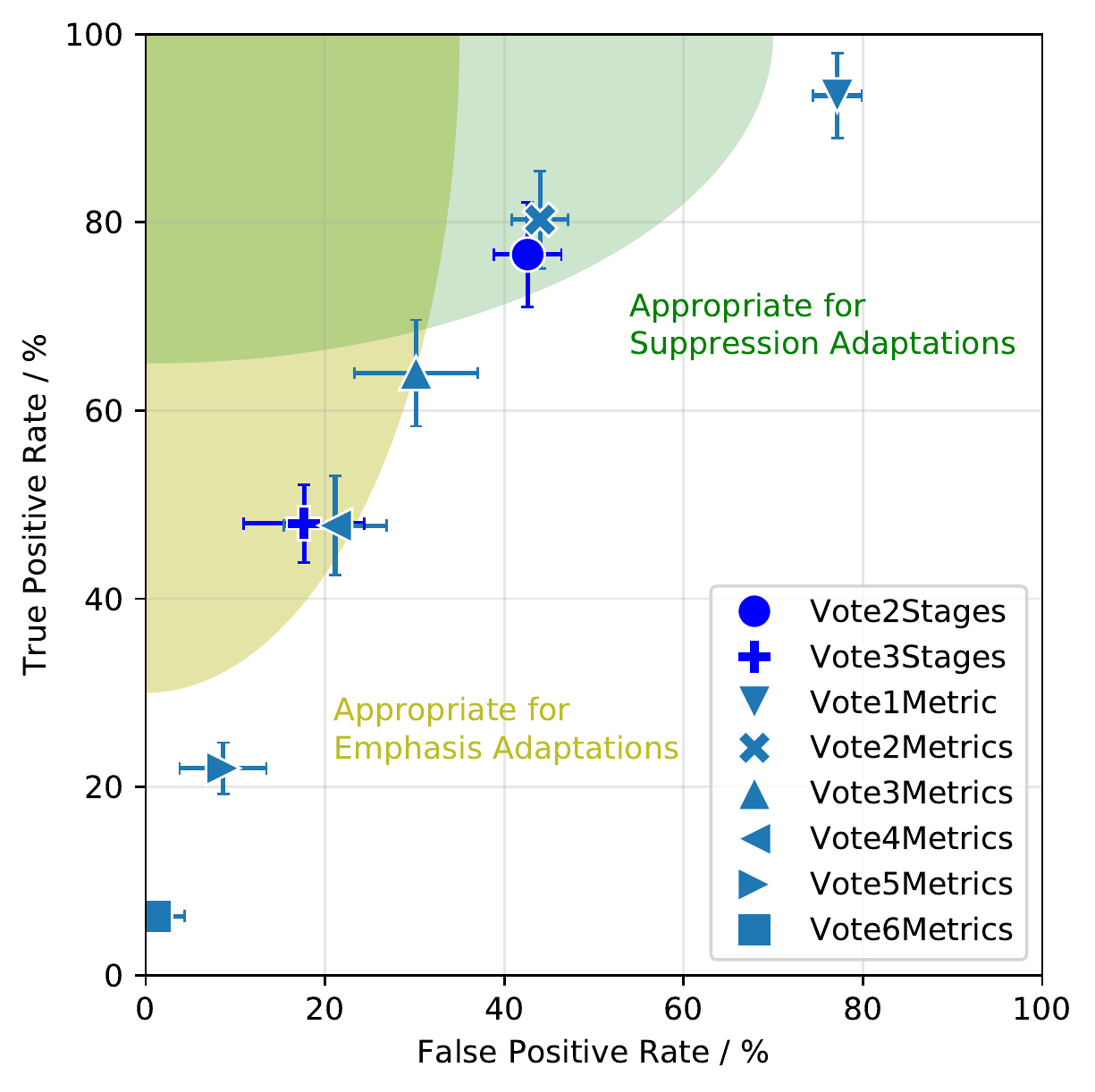}
\vskip -2mm
\caption{Variations in true positive versus false positive rate when changing the number of votes required to detect an AOI as relevant.
The error bars represent standard error over the 12 participants.
}
\label{fig:voting_thresholds}
\end{figure}

\begin{figure*}[t]
    \begin{subfigure}[t]{0.33\textwidth}
        \includegraphics[width=\columnwidth]{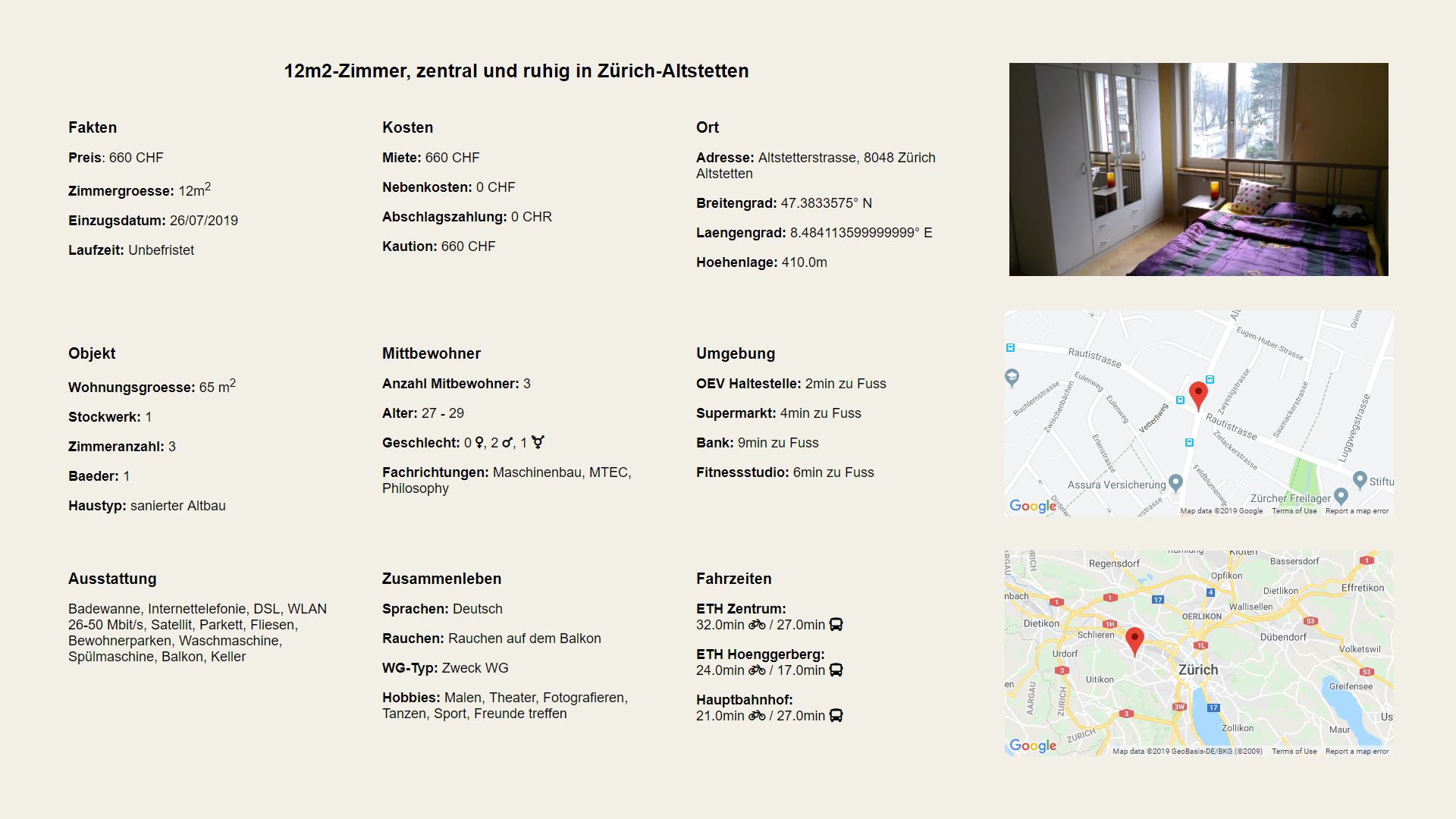}
        \vskip -1mm
        \caption{Default advertisement
        \label{fig:sample_default}
        }
    \end{subfigure}
    \hfill
    \begin{subfigure}[t]{0.33\textwidth}
        \includegraphics[width=\columnwidth]{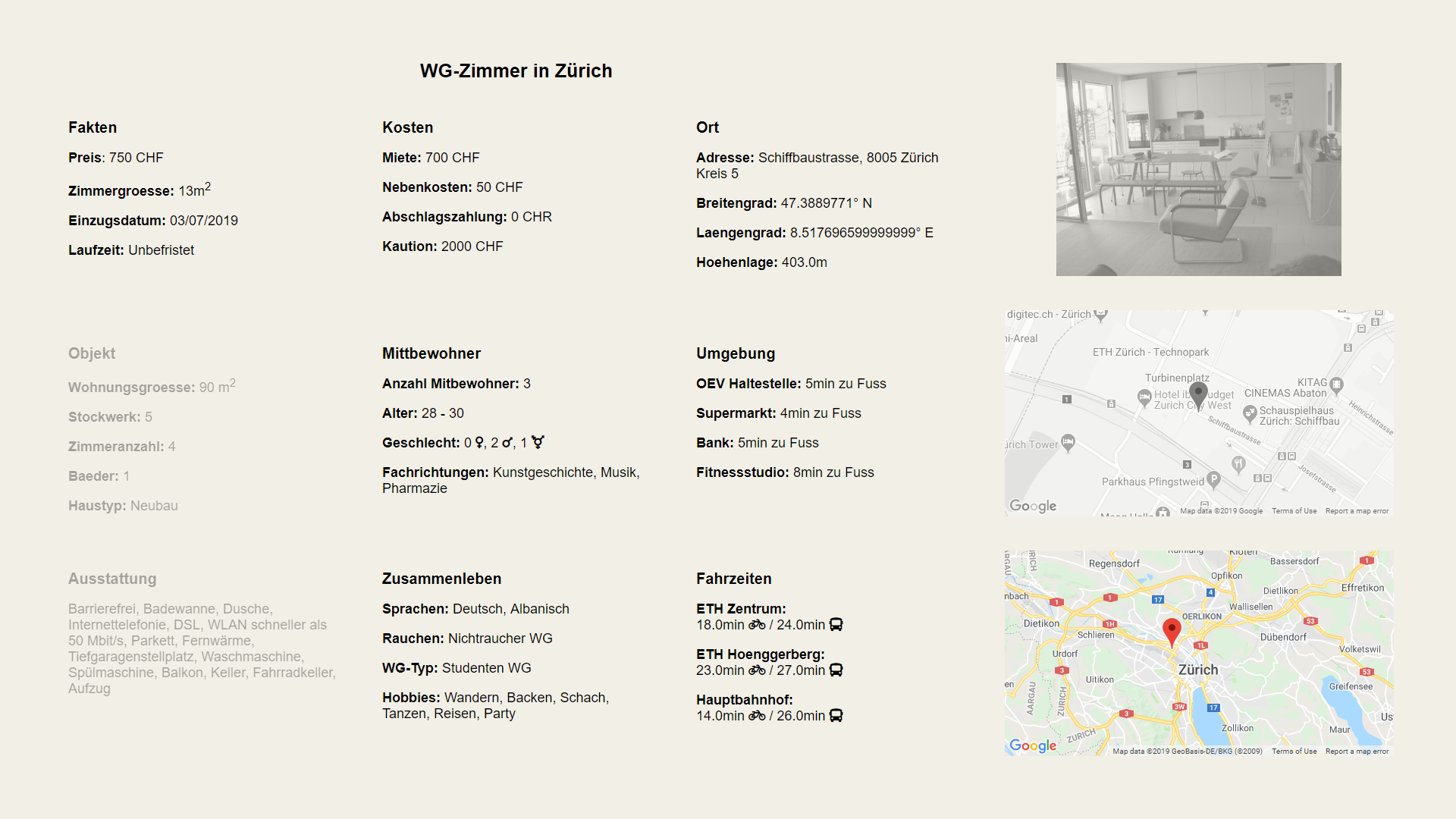}
        \vskip -1mm
        \caption{With fading adaptation (\twovotes)
        \label{fig:sample_fade}
        }
    \end{subfigure}
    \hfill
    \begin{subfigure}[t]{0.33\textwidth}
        \includegraphics[width=\columnwidth]{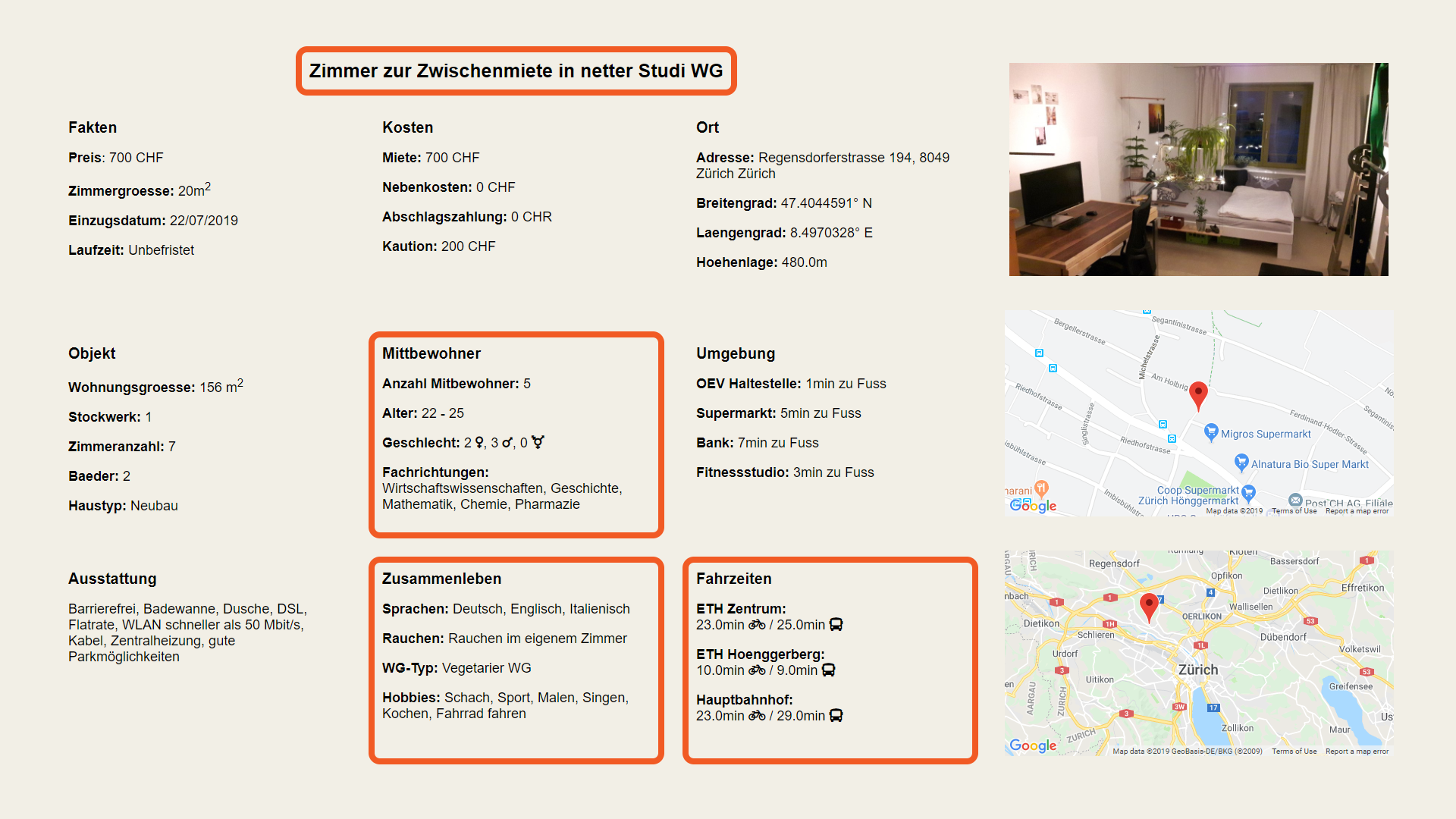}
        \vskip -1mm
        \caption{With highlighting adaptation (\threestages)
        \label{fig:sample_highlight}
        }
    \end{subfigure}
    \vskip -2mm
    \caption{Example advertisements from our room search application shown (a) without any adaption, (b) with fading, and (c) highlighting. }
    \label{fig:adaptations}
\end{figure*}

\subsection{Effect of vote threshold}
We collected a pilot dataset of gaze behavior during decision-making. This allowed us to analyze the performance of our voting approach and explore the effect of varying the number of metrics required to cast a vote in order for an AOI to be detected as relevant.

15 participants (13 male, 2 female) were shown a financial trading interface with information about a specific stock. Due to technical problems we excluded three participants. All of the remaining 12 participants had practical experience in stock trading (3 professionally employed, 9 students related fields).  For each participant, we recorded the eye gaze using a Tobii Pro Spectrum with a frequency of 150 Hz.
Information about the stock was organized within 12 AOIs. We based the interface design on existing trading interfaces and interviews we conducted with financial traders at a local branch of a global investment bank (see Figure \ref{fig:financialDashboard} in Appendix). Participants were asked to explore the data and state whether they would invest in the stock. After announcing their decision, we gave them a printout of the application and asked them to circle the parts of the interface (AOIs) they found to be most relevant to their task.

For each AOI, we established if a vote was cast by any of the six gaze metrics.
Considering the self-reports as ground truth labels, we can evaluate the performance of our voting approach
and explore the effect of different voting thresholds.
In order to determine the relevance of an AOI, we can require a minimum of $1$--$6$ gaze metrics to cast a vote. We can be more selective by considering votes from metrics associated to the same stage as redundant and require votes to come from metrics of $2$ or all $3$ stages.
Figure~\ref{fig:voting_thresholds} shows a clear trade-off between the true positive rate (relevant AOIs correctly detected) and the false positive rate (irrelevant AOIs detected as relevant) depending on the number of required votes. Shaded areas mark a true and false positive rate that seem acceptable when emphasizing or suppressing information.

We conclude that a good trade-off for emphasizing relevant information is achieved when requiring a minimum of $3$ votes each from a different stage (hereafter called \texttt{Vote3Stages}). Here, this yields a low false positive rate (18\%) reducing the risk to induce cognitive dissonance by emphasizing irrelevant information, while still detecting 48\% of relevant elements.
In the case of suppressing irrelevant information, a threshold of $2$ votes (hereafter called \texttt{Vote2Metrics}) yields a high true positive rate (80\%), reducing the risk of suppressing relevant information which would result in a higher interaction cost. The false positive rate of 44\% is acceptable, given that this is achieved after observing only one decision and without any explicit user input.

\section{Application to UI adaptation}
We developed a proof-of-concept application that visually adapts the interface using the \texttt{Vote2Metrics} and \texttt{Vote3Stages} recognizers. We chose a more commonly encountered scenario where users are faced with a decision whether to apply for an advertised apartment. This allows for natural and diverse personal preferences to exist. Indeed, a brief survey among university students indicated that while most found information such as price, location, and contract duration to be highly important, the relevance of information on potential cohabitants, amenities, or nearby landmarks varied.

To maximize the external validity of our findings, we carefully designed a realistic and useful interface in terms of content, layout, and adaptation method. The application shows individual room advertisements with information structured into 13 AOIs (including title), telling about a the room, its price and location, roommates, and additional details
such as estimated commute time to the university (see Figure ~\ref{fig:sample_default}). These were scraped from public apartment share websites, generated via the Google Maps API, or synthesized.
We implement two UI adaptation schemes with the goal to guide users' attention without being obtrusive. The application can \emph{suppress} irrelevant AOIs as detected by \texttt{Vote2Metrics} through reducing their opacity to 0.5. Thus, they blend into the background while still being legible (see Figure~\ref{fig:sample_fade}). Image-based AOIs such as embedded maps and photos are made to be monochrome to further reduce their visual saliency. The application can \emph{highlight} relevant AOIs as detected by \texttt{Vote3Stages} by drawing thick-bordered red boxes around them (see Figure~\ref{fig:sample_highlight}).

\subsection{User study}
We assessed the performance of our relevance detectors and their usefulness for UI adaptation in a lab study where we asked  participants to use our adaptive room search application to decide whether they would apply to a displayed advertisement.

\subsubsection{Participants}
Twenty participants (9 male, 11 female) were recruited through university mailing lists. The majority of them were students from various backgrounds. The average age was 26.2 years (SD = 4.6). The requirement for participation was having experience in searching for a room in a shared apartment and having a reasonable monetary budget.
One participant had to be excluded for being beyond the budget (= 1700 Swiss Francs) that our application could support in displaying realistic advertisements. All the other participants (N=19) had searched for a room in a shared apartment in the past 3.73 months on average (SD = 2.93) and had a mean budget of 810.50 Swiss Francs (SD = 225.31). 10 participants had vision corrections (6 with glasses, 4 with contact lenses)

\subsubsection{Setup \& experimental design}
The task consisted of a repetitive binary decision. Participants examined 35 room advertisements
and decided for each whether they would apply for the particular room. After announcing a decision they were asked about their confidence on a 7-point Likert scale.
Advertisements were shown in one of three conditions: \texttt{default} (no adaptation), \texttt{highlighting} and \texttt{fading}, referring to the adaptations described above and shown in Figure~\ref{fig:adaptations}.
We used a within-subjects design and randomized the order of conditions as described below.
We used a Tobii Pro Spectrum screen-based eye-tracker (with chin rest) recording at 150~Hz with a 24-inch monitor.

\subsubsection{Procedure}
The user study was conducted in German. Participants first gave their informed consent for data collection. Before starting the task they were asked to describe the criteria they consider important when searching for a room in a shared apartment. This was done for participants to form a mental model of their preferences beforehand.
After that, we showed them a randomly generated set of 35 room advertisements, none of which were more than 10\% above their stated budget.
During the first 4 advertisements, participants could familiarize with the interface. Eye-tracking data from the subsequent 10 advertisements (without any adaptation) were used to detect the relevance of each AOI. The last 21 advertisements were split into three blocks, each displaying 7 advertisements in one of the three conditions, where the order of conditions was randomized. After every advertisement, the user was asked to state whether they would apply for this room.
The person's eye gaze was recorded while the advertisements were displayed. The eye-tracker was calibrated using the built-in 5-point calibration procedure at the beginning of the study and between every block. The recorded gaze data was of high quality with a low calibration error ($M=0.14^{\circ}$) and high precision ($SD = 0.23^{\circ}$, $RMS = 0.21^{\circ}$).
The user could control the beginning (onset) and the end of an advertisement display by pressing the space bar.
At the end of the experimental session, participants filled a short questionnaire and went trough a semi-structured interview. There, we asked them what information they perceived as relevant for making a good decision and used the self-reported AOIs as ground-truth labels.
A session took 60 minutes on average and participants were compensated with 25 Swiss Francs for their participation.

  \begin{table}[t]
  \sffamily
  \small
    \captionsetup{font=small}
  \caption{Average true and false positive rates for recognizing relevance of AOIs. The chosen thresholds achieve different trade-offs and perform better than single metrics.}
  \label{tab:fpRates}
  \vskip -2mm
  \begin{tabular}{rll}

  \textbf{Metric} & T\textbf{P rate (SD)} & \textbf{FP rate (SD)}            \\ \hline
 \texttt{Vote2Metrics} & \textbf{97\% (7)} & 42\% (12) \\
 \texttt{Vote3Stages} & 80\% (19)& \textbf{17\% (11)} \\
  TFF    & 54\% (18)& 44\% (15) \\
  FPG    & 79\% (24)& 31\% (9) \\
  SPG    & 74\% (20)& 32\% (7) \\
  RFX    & 79\% (19)& 27\% (14) \\
  SFX    & 89\% (15)& 28\% (12) \\
  ODT    & 87\% (15)& 26\% (13) \\ \hline
  \end{tabular}
  \end{table}
\subsection{Results}

The room-search scenario and application was well perceived by all participants. When asked about realism, all users stated that the ads were genuine and comparable to those found on existing websites. One participant even asked if our application would be available for personal use. The ratio of yes/no answers on whether participants would apply for rooms in shown advertisements were 44\% and 56\% respectively.
Considering the complexity in providing both a balanced and realistic set of advertisements which do not cause cognitive dissonance due to small imprecision in detail, we achieved our goal of a realistic study environment and can expect our conclusions to generalize to real-world scenarios.

\subsubsection{Relevance detection}

On average, 97\% of relevant AOIs (as reported by participants) were recognized by  \texttt{Vote2Metrics} (on average 4.21 AOIs), thus only 3\%  of relevant information was wrongly faded, on average 0.2 AOIS,  minimizing the risk of cognitive dissonance. This comes at the cost of a false positive rate of 42\% (on average 3.63 AOIs) which were not faded out although they were not denoted as relevant. However, this does not induce any usability issues.
In the \texttt{highlighting} condition, the false positive rate could be reduced to 17\% by using \texttt{Vote3Stages}. This ensured that on average only 1.58 AOIs were highlighted although they were not regarded as relevant. As a trade-off the average true positive rate reduced to 80\%, resulting in 3.42 AOIs highlighted by red boxes. Such a trade-off is desirable since highlighting more than half of the AOIs would not have the desired effect.

Table~\ref{tab:fpRates} compares the performance of these recognizers against that of individual metrics. While overall dwell time (ODT) and sum of fixations (SFX) perform relatively well in comparison to others, \texttt{Vote2Metrics} and \texttt{Vote3Stages} better maximize true positives and minimize false positives respectively, as required for our adaptations. In particular, the voting scheme can recognize edge cases where e.g. a relevant AOI is characterized by a short time to first fixation (TFF) and several refixations (RFX) but not a high ODT.

\begin{figure}[t]
\centering
\begin{subfigure}{0.49\columnwidth}
  \includegraphics[width=\columnwidth]{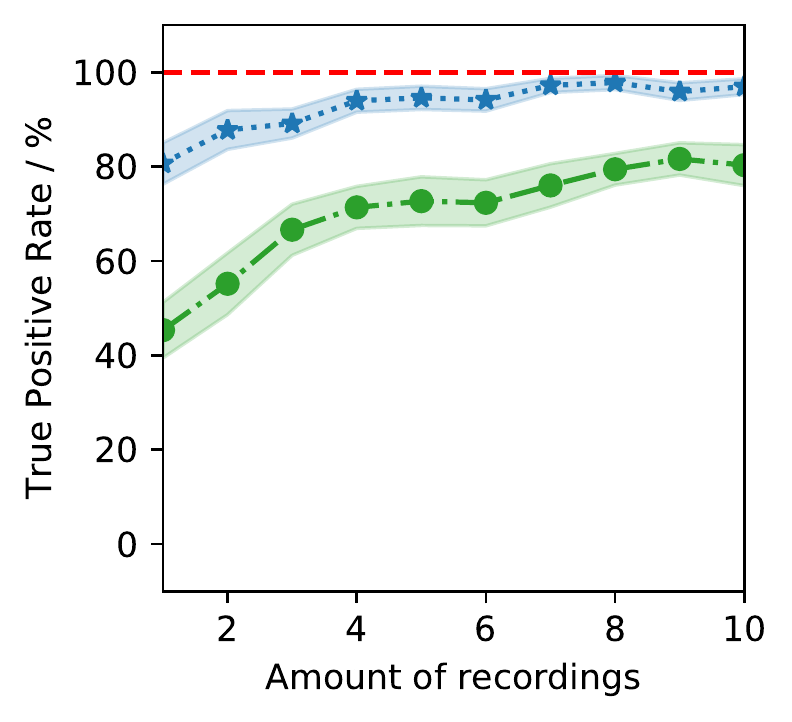}
  \vskip -2mm
\end{subfigure}
\hfill
\begin{subfigure}{0.49\columnwidth}
  \includegraphics[width=\columnwidth]{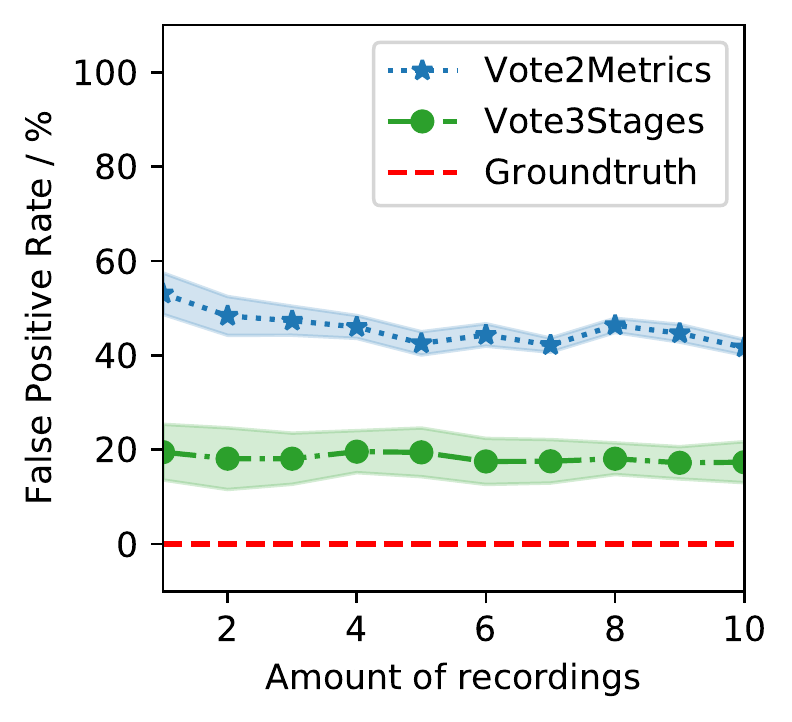}
  \vskip -2mm
\end{subfigure}
\caption{Increase in true positive rate and decrease in false positive rate with larger number of recordings (shaded is standard error).
}
\label{fig:success_over_time}
\end{figure}

In our pilot evaluation, relevance detection was based on gaze data of only one decision. In contrast,  in this study we recorded eye movements over 10 subsequent decisions. As shown in Figure \ref{fig:success_over_time}, this has a positive effect on the true positive and false positive rates of our two recognizers. Still, already after one observation, we observe a high true positive rate of over 80\% with \twovotes. This is consistent with the pilot data. The false positive rate with  \threestages is consistently low indicating its robustness even with one or few observations. With more recordings, effects from the specific content of particular advertisements can be smoothed out, and a better understanding of task and user relevance can be made. Consequently, the standard error of our approach decreases.

\begin{figure*}[t]
    \begin{subfigure}{0.33\textwidth}
        \includegraphics[width=\columnwidth]{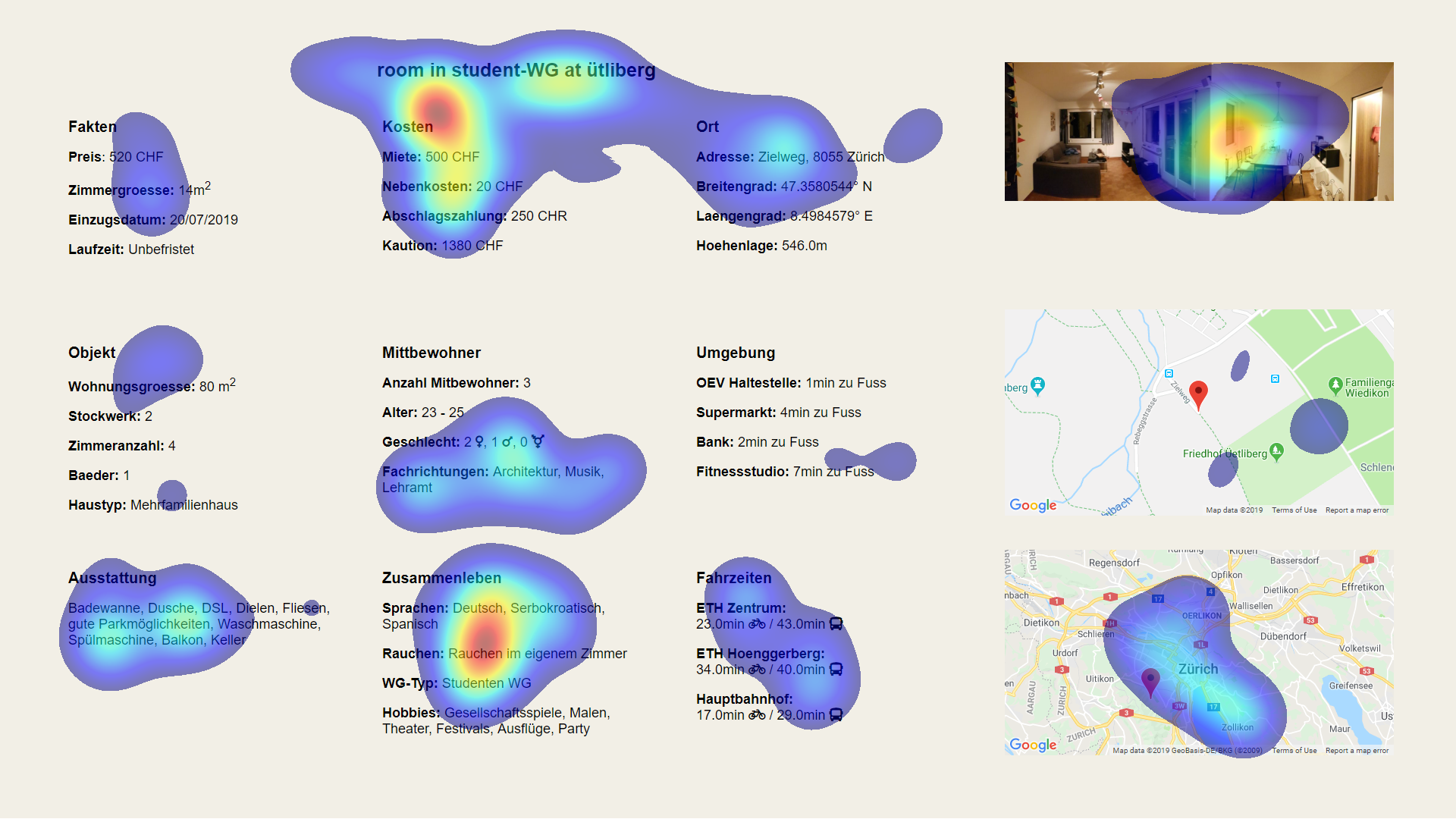}
        \vskip -1mm
        \caption{Default Condition}
    \end{subfigure}
    \hfill
    \begin{subfigure}{0.33\textwidth}
        \includegraphics[width=\columnwidth]{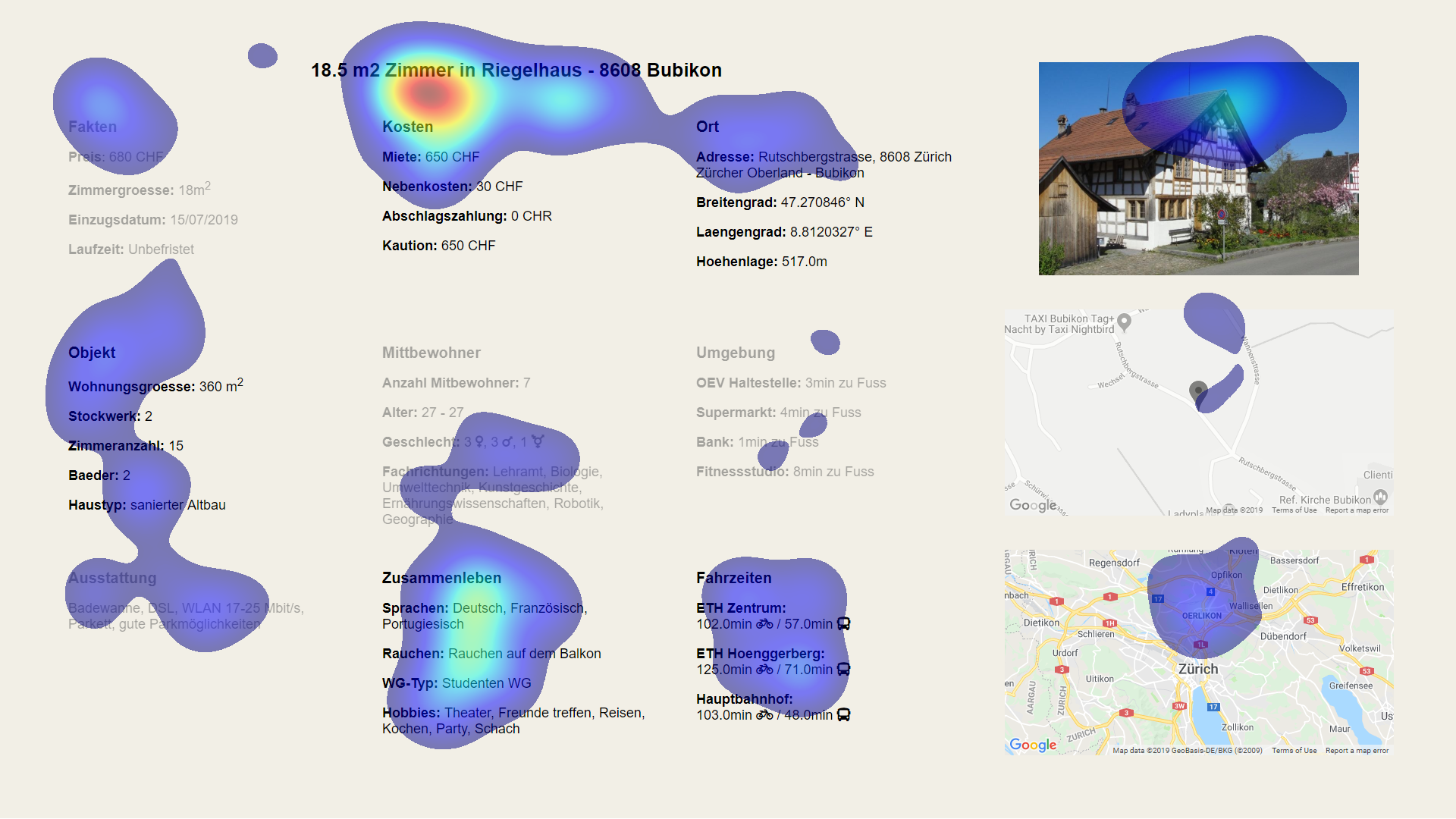}
        \vskip -1mm
        \caption{Fading Condition (\twovotes)}
    \end{subfigure}
    \hfill
    \begin{subfigure}{0.33\textwidth}
        \includegraphics[width=\columnwidth]{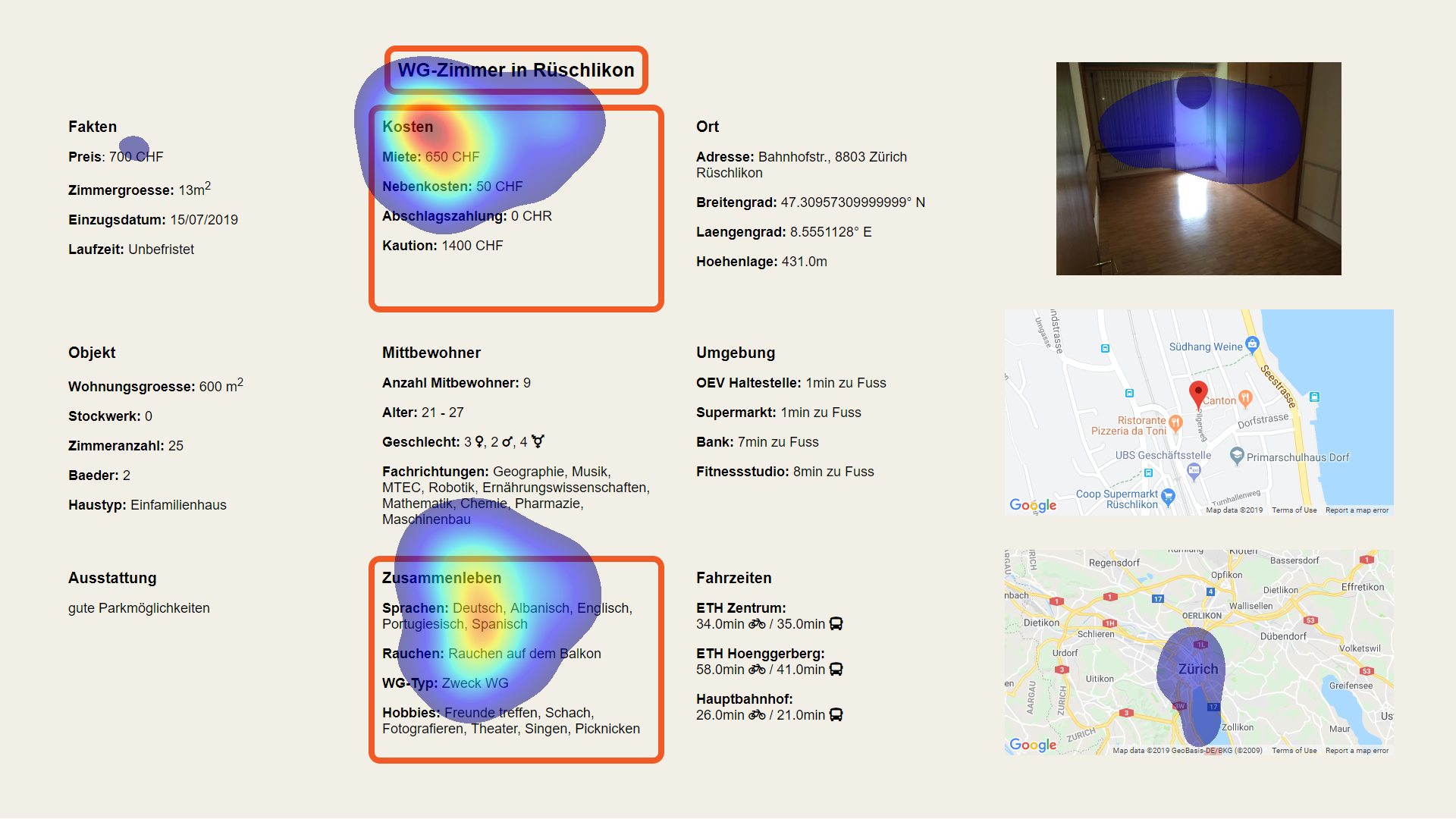}
        \vskip -1mm
        \caption{Highlighting Condition (\threestages)}
    \end{subfigure}
    \vskip -1mm
    \caption{Accumulated fixations over 7 trials per experimental condition, for participant 4. It can be seen that both adapted conditions (highlighting and fading) yield better-concentrated fixation clusters compared to the unadapted default condition.}
    \label{fig:heatmap_conditions}
\end{figure*}

\begin{table}[b]
\sffamily
\small
 \captionsetup{font=small}
\caption{Overview of the dependent variables measured in the three study conditions.}
\label{tab:measured_effects}
\vskip -3mm
\begin{tabular}{lllll}

\textbf{Metric  }            & \textbf{\texttt{Default}}  & \textbf{\texttt{Fading}} & \textbf{\texttt{Highl.}} & \textbf{Note} \\
\hline
Task execution      & 20.22s  & 20.8s  & 19.9s  &                     \\\vspace{3pt}
time (SD)           & (6.12)  & (7.87) & (6.54) &                     \\
Confidence          & 5.55    & 5.6    & 5.61   & 1 (not confident) --\\\vspace{3pt}
(SD)                & (0.5)   & (0.74) & (0.66) & 7 (very confident)  \\
Perceived inform.   & 4.53    & 4.68   & 4.9    & 1 (overwhelming) -- \\\vspace{3pt}
density (SD)        & (1.61)  & (0.95) & (1.52) & 7 (balanced)        \\
Adapt. facilitates          & -       & 3.21   & 4.63   & 1 (never) --        \\\vspace{3pt}
 task (SD)    & -       & (1.51) & (2.11) & 7 (always)          \\
Preference count    & 8       & 4      & 6      & 1 undecided         \\
\hline
\end{tabular}
\end{table}

Participants commented positively on the effectiveness of the relevance detector, stating for example that \emph{``Nothing was faded out, that I actually wanted to read.''}, \emph{``The red boxes highlight the most important information''}, and
\emph{``First I felt guilty that I did not look at all the information, but when elements were faded out, [...] I was okay with not looking at all of the information.''}
Only one participant said that an AOI they considered important was not highlighted.
This concerned the AOI including the net rent, which the participant memorized at a glance and did not have to return to, thus avoiding the triggering of votes from any single gaze metric.

\subsubsection{Effect of UI adaptation}

Overall, we could not observe any change in task performance as an effect of the UI adaptation (Table \ref{tab:metrics_list}). Task execution time (TET) remained similar throughout all conditions. Participants reported a slight increase in the confidence of their decision during either one of the adapted conditions. We statistically tested the differences between conditions using Linear Mixed-Effects Models with a random part for participant-level effects (using the LME4 package for R and the lmerTest package to calculate p-values via Satterthwaite's degrees of freedom method). No significant difference was found for neither TET nor confidence ($p>.05$ in all cases).
Nevertheless, when asked about perceived information density on a 7-point Likert scale, a small, yet positive trend was observed for the adapted conditions. These results were similar to answers on whether the adaptation made the task easier. All measurements can be found in Table \ref{tab:measured_effects}.

When asked which version they would like to use, 10 / 18 participants stated that they preferred one of our adapted interfaces over the default case (one was undecided).
Some participants preferred \texttt{highlighting}, stating for example that \emph{``The interface had a better structure with red boxes and made it easier to search for information''}.
However, others disagreed saying:
\emph{``I prefer fading because default showed too much information and it was easier when things were faded out. On the other hand, I did not like coloring, because it was weird.''}
This indicates that a visual interface adaptation scheme should consider users' preferences and offer multiple options.

Interestingly, several participants that preferred the default version did so due to concerns about their behavior being influenced by an algorithmic system, saying: \emph{``I like to decide for myself what is relevant and what not.''} and \emph{``In [the] default [condition,] information is weighted equally. I do not expect from an application to tell me what is relevant, I want to have my freedom of decision.''}.

\section{Discussion}
We proposed an approach to detect the relevance of information displayed by a UI during decision-making from user gaze behavior. Informed by the eye-tracking and decision-making literature, we carefully selected a set of six gaze metrics which each capture different aspects of people's gaze behavior.
By combining them in a voting scheme we can obtain a more holistic view of a person's attention during the decision process to infer the relevance of the displayed information.
At the same time, this makes the detection robust against variations in the gaze behavior across users due to differences in decision strategies.
Indeed, our voting schemes (\twovotes and \threestages) performed decidedly better compared to the individual metrics (cf. Table~\ref{tab:fpRates}).
Although there are several more gaze metrics used by prior work such as duration of last fixation, inter-saccadic jumps or Krejtz's coefficient $\kappa$ \cite{krejtz2017}, we limited our approach to the most commonly used ones that covered gaze characteristics during the different stages of decision-making. This also ensures that our relevance detector is not tuned to a specific task but applicable to very different domains, as shown in this paper.

\subsection{Generalizability of the approach}

We evaluated our approach with data from two applications. Both had a static graphical layout and presented the user with a yes / no decision.
The latter is not a requirement for the applicability of the voting scheme. The three stages of decision-making which form the theoretical basis of our approach were associated also with other types of decisions, such as when choosing between different alternatives~\cite{russo1994}.
However, future works needs to assess the applicability of our approach for dynamic interfaces. In such a case, the voting scheme could be changed to take into account a longer or re-occuring orientation phase of the user for example by reducing the weight of votes casted by metrics associated with the first stage.
We organized the presented information into well-structured AOIs to ensure reliable gaze data. As eye tracking technology improves, the same approach can be used to detect information relevance for individual pieces of information also in less structured interfaces.

\subsection{Trading-off true and false positives}
For an adaptive UI to be successful, special care needs to be taken in order to avoid wrong adaptations that could increase the interaction cost or induce cognitive dissonance~\cite{Findlater2009}. Accordingly, in this paper, we have argued that depending on whether a UI emphasizes relevant content or suppresses irrelevant information, a recognizer should aim to minimize its false positive rate or focus on maximizing true positives.
We have shown that by varying the minimum number of required votes, the trade-off between true and false positives can be easily adjusted in a predictable way. Figure~\ref{fig:cross-dataset} shows that this is the case in both the stock-trading as well as the room-search dataset, indicating that this property generalize across different task domains and holds from the first observed decision, as well as for ten observations.
Our work enables UI designers to easily determine the voting scheme that yields the best trade-off for their specific application.

\begin{table}[t]
\sffamily
\small
\captionsetup{font=small}
\vskip -3mm
\caption{Performance comparison of our voting approach to standard machine learning methods trained on the room search data. Hyperparameters were tuned on the pilot data.}
\label{tab:ours_vs_machine_learning}
\vskip -2mm
\begin{tabular}{llll}

\textbf{Model}                                       & \textbf{TP rate (SD)} & \textbf{FP rate (SD)} & \textbf{F1 score (SD)} \\
\hline
\texttt{Vote3Stages}                        & 80\% (19)    & 17\% (11)    & \textbf{0.72} (0.16) \\
\texttt{Vote2Metrics}                  & \textbf{97\%} (7) & 42\% (12)    & 0.68 (0.13) \\
\hline
SVM                                         & 68\% (23)    & 15\% (13)    & 0.64 (0.18) \\[-1mm]
{\scriptsize(Linear kernel, C = 0.1)} & & \\
Decision Tree                               & 61\% (20)    & 18\% (12)    & 0.59 (0.20) \\
MLP                                         & 64\% (41)    & 20\% (16)    & 0.49 (0.32) \\[-1mm]
{\scriptsize(ReLU, h = 16, lr = 0.01)} & & \\
Log. Regression                             & 61\% (28)  & \textbf{14\%} (12) & 0.58 (0.23) \\ [-1mm]
{\scriptsize(Elastic net reg., C = 1.0)} & & \\
\hline
\end{tabular}
\end{table} \subsection{Effectiveness of the UI adaptation}
The large majority of users confirmed the efficacy of our room search application in detecting relevant information and many preferred the adapted version. By using different recognizers for highlighting and fading we could minimize the risk of cognitive dissonances or usability issues due to wrong adaptations.
Nevertheless, on average we did not find a significant impact on task execution time, confidence in decision-making or perceived information load.
At the same time, participants showed large individual differences. For example, we could observe individual cases where the adaptation guided the eye gaze of the participant, allowing them to better focus their information retrieval process, as shown for an exemplary participant in Figure~\ref{fig:heatmap_conditions}.
A possible explanation could be that the chosen adaptations (red boxes and reduced opacity) are not effective in facilitating the decision process in the first place. The few studies that assess the effect of visual adaptations of graphical UIs are often contradictory and do not consider a large range of adaptation techniques~\cite{Deuschel2016}. While this paper offers a method to detect information relevance, we need future work to establish better ways to effectively adapt UIs to facilitate the decision-making process. Such work should take into account the concern of some users that they are manipulated by the application and therefore preferred the default and non-adapted version. Similar concerns were observed by  by prior work~\cite{Yang2014CHI,Park2018CHI}. Adaptive interfaces could offer explicit control mechanisms for users to customize adaptations or to review and change the inferences made about them.

\subsection{Relation to supervised classification}
While prior work introduced no directly comparable methods, machine learning methods such as logistic regression have been employed in related areas of visual target search and image relevance\cite{klami2010} .
Unlike our method, such learning-based methods require ground-truth data from the exact same task and setting.
For a fair comparison in the case where the task is known, we performed a leave-one-out cross-validation on our room-search data for several machine learning methods.
Table~\ref{tab:ours_vs_machine_learning} summarizes the results and compares them against \twovotes and \threestages.
It shows a clear dominance in true positive rates for both our methods, with \threestages having a  comparably low false positive rate.
We also find that our methods score well in terms of F1 score -- a metric often used for evaluating binary classifier systems.

\begin{figure}[t]

    \centering
    \begin{subfigure}{0.49\columnwidth}
        \includegraphics[width=\columnwidth]{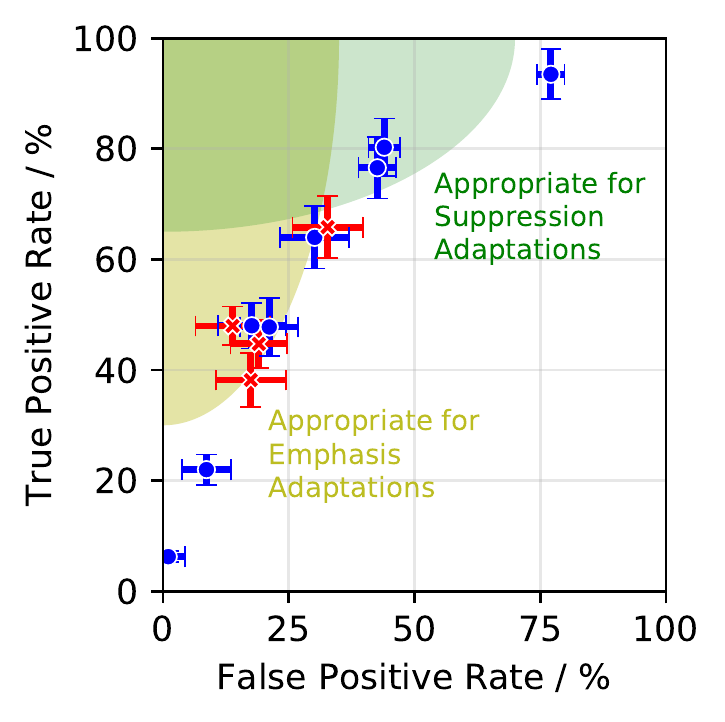}
        \vskip -2mm
        \caption{Trained on room search data, evaluated on pilot data}
        \label{fig:rs_to_fd}
    \end{subfigure}
    \hfill
    \begin{subfigure}{0.49\columnwidth}
        \includegraphics[width=\columnwidth]{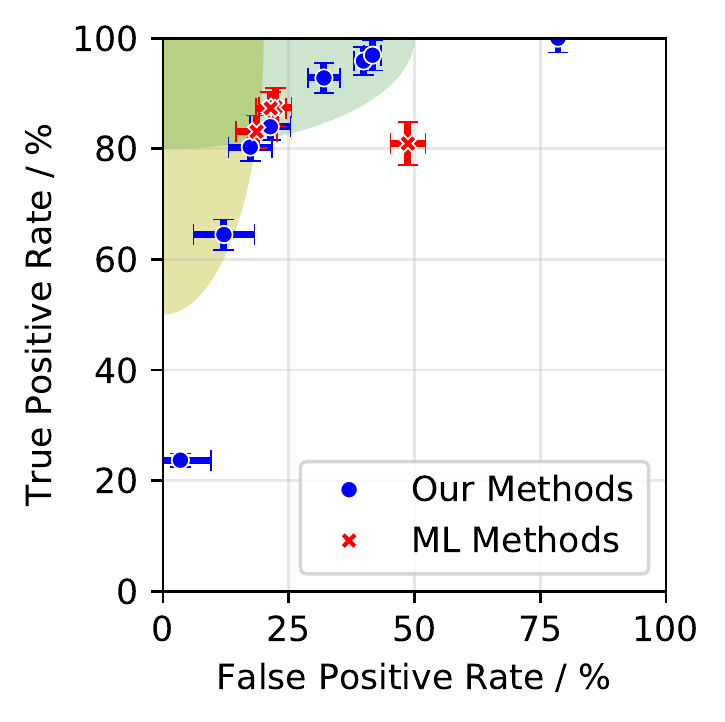}
        \vskip -2mm
        \caption{Trained on pilot data, evaluated on room search data}
        \label{fig:fd_to_rs}
    \end{subfigure}
    \caption{
        Performance of standard ML methods in comparison to our training-free voting approach. Voting yields a predictable trade-off between true and false positives across tasks which cannot be achieved with standard ML methods.
    }

    \label{fig:cross-dataset}
\end{figure}

More interesting, however, is to see how these methods generalize to new tasks and settings. We thus performed two experiments: (a) one where we train the machine learning methods on the room search data (10 decisions), and evaluate on the stock trading data (1 decision), and (b) the other way around.
As seen in Figure~\ref{fig:cross-dataset}, our approach shows the predictable trade-off through varying the number of votes in both task domains, whereas the machine learning methods have no clear consistency in performance, all achieving low false positive rates in one and high true positive rates in the other case.
This would make it challenging for a UI designers to anticipate how their system would perform in practise, whereas our method enables them to fit the system's behavior to the chosen adaptation scheme by selecting an appropriate threshold.

\section{Conclusion}
This paper contributes an approach for detecting the relevance of information displayed by a UI  during the decision-making process of a user. This could enable new adaptive applications that change their interface in response to the interest of a specific user to increase the user's experience and facilitate her decision-making.
Based on two empirical datasets, we show that our approach can trade-off true and false positive rates in a manner where performance changes can be anticipated. It does this without requiring any explicit feedback from users by only observing their gaze behavior during the decision process. Therefore, we combine well-established metrics from the eye-tracking literature in a voting approach which ensure that the recognition is robust to the large variations in gaze behavior during the decision process and across users who might employ different decision strategies. In contrast to supervised learning approaches, it does not require any training or parameter-tuning and is simple to understand, implement and adjust. An open-source python implementation of our approach is available at \url{https://ait.ethz.ch/projects/2020/relevance-detection}

\begin{acks}
The authors thank Christoph Gebhardt for insightful discussions.
This project has received funding from the European Union’s Horizon 2020 research and innovation program / from the European Research Council under the Grant Agreement No. StG-2016-717054.
\includegraphics[width=0.4\columnwidth]{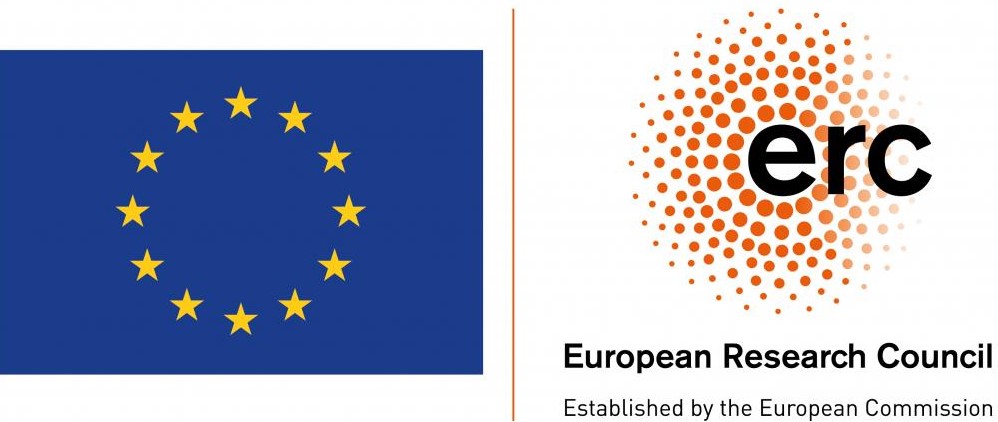}
\end{acks}
\balance{}

\bibliographystyle{ACM-Reference-Format}
\bibliography{references}

\newpage
\appendix
\section{Appendix}

\vspace{6pt}
\begin{minipage}[h!]{0.9\textwidth}
 The following image shows the financial trading interface used in the pilot study.  Participants were asked to evaluate a stock quote and make an investment decision. The displayed information is organized into 12 AOIs. The recorded gaze was used to test our voting approach in comparison to self-reported ground-truth.

  \includegraphics[width=\textwidth]{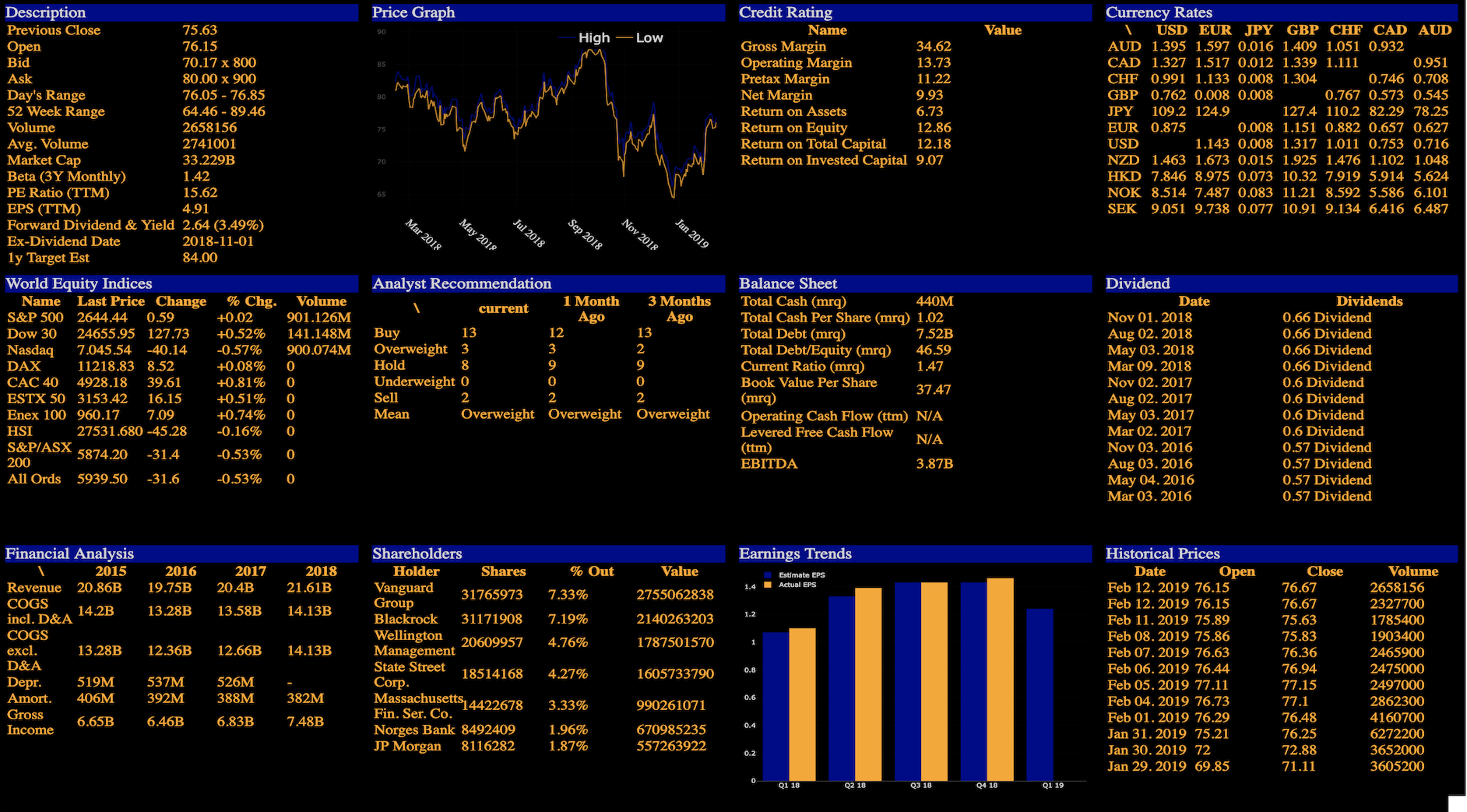} \label{fig:financialDashboard}

 \vspace{10pt}
 The following image shows a larger version of Figure~\ref{fig:adaptations}, the room search application. Similar to the financial dashboard above, information about a room is spatially grouped into 12 AOIs to ensure reliable gaze information.

 \includegraphics[width=\textwidth]{figures/exemplar_default_cond.png}
\end{minipage}

\begin{minipage}[h!]{\textwidth}

\end{minipage}

\end{document}